\documentclass[]{raa}            % referee version: for submission
\usepackage{graphicx,times}
\usepackage{amssymb}
\usepackage{natbib}
\begin{document}%%
   \title{Multi-band Photometric and spectroscopic analysis of the dwarf novae IU Leo}
\volnopage{ {\bf 0000} Vol.\ {\bf 0} No. {\bf XX}, 000--000}
\setcounter{page}{1}
\author{Y. H. Chen\inst{1,2}$^*$, \ C. M. Duan\inst{1,3}, \ and H. Shu\inst{1}}
\institute{\inst{1} Institute of Astrophysics, Chuxiong Normal University, Chuxiong 675000, China; {$yanhuichen1987@126.com$}\\
           \inst{2} International Centre of Supernovae (ICESUN), Yunnan Key Laboratory, Kunming 650216, China;\\
           \inst{3} Faculty of Science, Kunming University of Science and Technology, Kunming 650093, China\\
\vs \no
{\small Received [0000] [July] [day]; accepted [0000] [month] [day] }}

\abstract{IU Leo was first identified as a cataclysmic variable star in 2006. Based on an image data and a distance value, we derived that the circumbinary envelope of IU Leo was $\sim$3,745\,AU on the optical band. According the multi-band photometric data, we calculated a $T_{eff}$ of a few hundred Kelvin for the circumbinary envelope of IU Leo. We reviewed the physical parameters of IU Leo and simulated the evolution process using a stellar evolution code MESA with $M_{1}$=0.982\,$M_{\bigodot}$, $M_{2}$=0.835\,$M_{\bigodot}$, and an orbital period of 0.376308\,days. The evolved other parameters are basically consistent with the parameters in the literatures. Based on the quiescence Kepler Mission 2.0 light curve, the quiescence Transiting Exoplanet Survey Satellite light curve, and 89 Large Sky Area Multi-Object Fiber Spectroscopic Telescope medium resolution spectra, we derived an orbital period of 0.376307 $\pm$ 0.000004\,days, 0.3762 $\pm$ 0.0001\,days, and 0.3763\,days for IU Leo respectively. These orbital periods are basically consistent with the results of previous studies. According to light curve of IU Leo from American Association of Variable Star Observers, we reported three new outburst spectra from the Large Sky Area Multi-Object Fiber Spectroscopic Telescope low resolution catalogue with part Balmer emission lines overlap on their absorption lines. Many H, He, C, N, O, Na, Mg, Si, and Ca neutral and ionized lines are identified, which are produced by different mechanisms. In the future, we will conduct more comprehensive and in-depth research on CVs based on multi-band photometric and spectroscopic data.
 }

\authorrunning{Y. H. Chen, C. M. Duan, \& H. Shu}       %author_head in even pages
\titlerunning{Multi-band analysis of dwarf novae IU Leo}  % title_head in odd pages
\maketitle

\section{Introduction}

Stars constitute the primary objects among celestial bodies, and stellar physics serves as the foundation of astrophysics. Therefore, the study of stellar structure and evolution holds universal significance. From both the perspective of birth rate simulation (Machida et al. 2005) and observational statistics (Sana et al. 2013, El-Badry \& Rix 2018), binaries account for a significant proportion of stars. Based on spectra released by the Large Sky Area Multi-Object Fiber Spectroscopic Telescope (LAMOST, Zhao et al. 2012, Cui et al. 2012), Qian et al. (2018) studied 2020 EA-type eclipsing binaries and investigated their physical properties and evolutionary states. Based on photometric database of the Kepler Mission (Koch et al. 2010), Li et al. (2020) derived physical parameters of 380 EW-type eclipsing binaries and analyzed the relationships between them. The research on binaries can accurately obtain stellar masses and radii. Xiong et al. (2023) reported a catalog of 184 binaries with accurate masses, accurate radii, and independent atmospheric parameters. Black hole binary (Abbott et al. 2016) and neutron star binary (Abbott et al. 2017) are important sources and key detection targets of gravitational waves. Luo et al. (2025) reported a born super-Chandrasekhar candidate binary including an ultramassive white dwarf and a hot subdwarf. Binary systems are crucial celestial objects that contain abundant physical laws.

Studying accreting white dwarfs is of great significance for studying matter transfer and accretion physics. Webb (2023) summarized the classification of accreting white dwarf binary stars, including white dwarf companion stars (AM CVn), main sequence companion stars (ultra soft X-ray sources and cataclysmic variables), and red giant companion stars (symbiotic stars). Cataclysmic variables (CVs) are an interesting subclass of accreting white dwarf binaries, where a white dwarf (the primary star) accretes matter from its companion (the secondary star) via Roche lobe overflow. Based on the Global Astrometric Interferometer for Astrophysics (GAIA, Gaia Collaboration et al. 2016) DR3, Canbay et al. (2023) established spatial distribution, galactic model parameters, and luminosity function for 1587 CVs. The Sloan Digital Sky Survey V (SDSS-V) (Kollmeier et al. 2017) contributed a dedicated survey for white dwarfs, single and in binaries, and Inight et al. (2025) reported an analysis of the spectroscopy of 504 CVs and CV candidates from SDSS-V. In 2021, Sun et al. studied a catalog of 323 CVs from LAMOST DR6. The study of CVs represents a key research focus in binary star astrophysics.

The CVs generally include classical novae, dwarf novae, novalikes, and magnetic CVs. Unlike the explosive burning of matter accreted by classical novae (Gallagher \& Starrfield 1978), the explosion of dwarf novae comes from the instability of the accretion disk (Cannizzo 1993). The instability within the accretion disk of dwarf novae exhibits typical dynamical timescales, which are more transient than those of thermonuclear reactions and contain faster physical processes. Conducting in-depth research on dwarf novae can provide key clues for understanding the accretion process and matter transfer mechanism under dynamic timescales, and help reveal the essential laws of rapid exchange of matter and energy in CVs. In 2006, Greaves first identified IU Leo (GSC 00847-01021) as a new brightish CV (dwarf novae) in Leo. Subsequently, IU Leo (RA = 10 57 56.29, Dec = +09 23 14.9) was observed by multiple telescopes on multiple observation missions. Ritter \& Kolb (2003) reported a catalogue (from June 30, 2003 to December 31, 2015) of binary stars, including CVs, and recorded IU Leo as a dwarf novae since December 31, 2007. Since June, 2008, the American Association of Variable Star Observers (AAVSO, Percy \& Mattei 1993) has carried out sustained observations of IU Leo year after year.

In this paper, we plan to conduct a systematic analysis of multi-band photometry and spectroscopy on IU Leo. In Sect. 2, a preliminary overall analysis and a simulation verification of IU Leo are performed. The observation data are from GAIA, SDSS, Two Micron All Sky Survey (2MASS, Skrutskie et al. 2006), Wide-Field Infrared Survey Explore (WISE, Wright et al. 2010), and so on. The simulation code is Modules for Experiments in Stellar Astrophysics (MESA, Paxton et al. 2015). In Sect. 3, we conduct a photometric analysis of IU Leo based on K2 (Kepler Mission 2.0), Transiting Exoplanet Survey Satellite (TESS, Ricker et al. 2014), and AAVSO. A spectroscopic analysis of IU Leo is carried out based on LAMOST and Hubble Space Telescope (HST, Shore 1992) in Sect. 4. At last, we give a discussion and conclusions in Sect. 5.

\section{A preliminary overall analysis and a simulation verification of IU Leo}

In Sect. 2, we first visually understand IU Leo through an image data, study the radiation of IU Leo through multi-band apparent magnitude data, and explore physical properties of circumbinary envelope of IU Leo through mid-infrared apparent magnitude. Then, we review the basic physical parameters of IU Leo in previous papers and conduct preliminary simulation calculations on IU Leo using MESA. We preliminarily evaluate the rationality of the basic physical parameters of IU Leo using the results of MESA evolution.

In Fig. 1, we show the image data for IU Leo from SDSS\_DR15. There is no corresponding spectrum for IU Leo in SDSS\_DR15. We can see that the circumbinary envelope around IU Leo spreads for about 5 arcseconds. The latest research (Canbay et al. 2023, Bruch 2024) shows that the distance of IU Leo is 749 $\pm$ 15 pc. Therefore, we can infer that the circumbinary envelope of IU Leo roughly extends to $\sim$3,745\,AU on the optical wavelength band. In Table 1, we show the apparent magnitude for IU Leo from SDSS, 2MASS, and WISE respectively. Basically, as the wavelength increases, the apparent magnitude gradually decreases. In the mid-infrared band of 22\,$\mu$m, the apparent magnitude is 7.947. According to Wien's displacement law, the peak wavelength of 22\,$\mu$m corresponds to a $T_{eff}$ of 132\,K. The scale of the circumbinary envelope is several thousand astronomical units on the optical wavelength band, and the $T_{eff}$ is a few hundred Kelvin on the mid-infrared band.

\begin{figure}
\begin{center}
\includegraphics[width=8.0cm,angle=0]{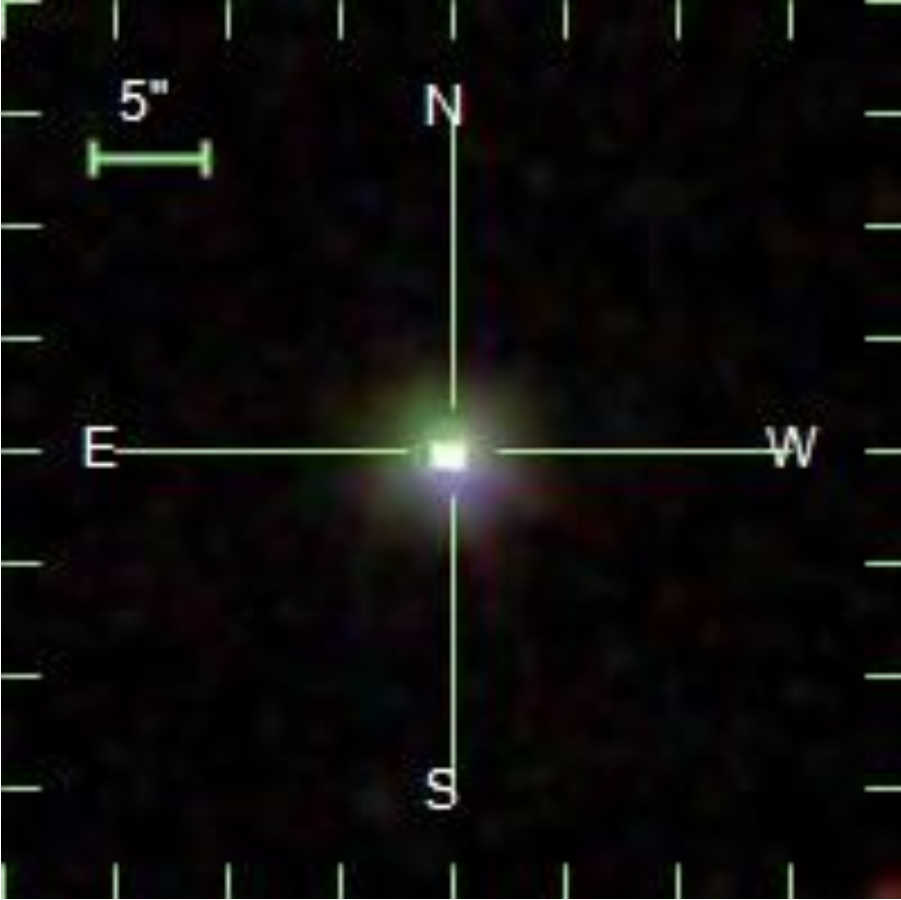}
\end{center}
\caption{The image data for IU Leo from SDSS\_DR15. The image was observed on March 12, 2002, corresponding to a MJD of 52,345.}
\end{figure}

\begin{table*}
\begin{center}
\caption{The apparent magnitude for IU Leo. The values of ugriz, JHK, and w1w2w3w4 are released by SDSS, 2MASS, and WISE respectively. The center wavelength of ugriz, JHK, and w1w2w3w4 are from Fukugita et al. 1996, Skrutskie et al. 2006, and Wright et al. 2010 respectively.}
\begin{tabular}{lcccccccccccccccccccccccccc}
\hline
filter             &u            &g          &r           &i            &z           &J           &H          &K         &w1        &w2        &w3         &w4        \\
$\lambda$($nm$)    &\,355.7      &\,482.5    &\,626.1     &\,767.2      &\,909.7     &\,1,250     &\,1,650    &\,2,160   &\,3,400   &\,4,600   &\,12,000   &\,22,000  \\
\hline
mag                &\,16.53      &\,15.88    &\,15.21     &\,14.92      &\,14.76     &\,13.720    &\,13.201   &\,13.057  &\,13.069  &\,13.003  &\,11.406   &\,7.947   \\
err                &\,0.01       &\,0.00     &\,0.00      &\,0.00       &\,0.00      &\,0.028     &\,0.034    &\,0.036   &\,0.012   &\,0.027   &           &          \\
\hline
\end{tabular}
\end{center}
\end{table*}

\begin{table}
\begin{center}
\caption{Table of physical parameters of IU Leo. The reference ID\_1, ID\_2, and ID\_3 are from Ritter \& Kolb 2003, Thorstensen et al. 2010, and Pala et al. (2017) respectively.}
\begin{tabular}{llllllllllll}
\hline
parameters                  &ID\_1                &ID\_2                  &ID\_3                \\
\hline
period(days)                &\,0.376308           &\,0.376308             &\,0.376306           \\
maximum(mag)                &\,12.9               &\,13.0                 &                     \\
minimum(mag)                &\,15.7               &\,15.5                 &                     \\
q=$M_{2}/M_{1}$             &                     &\,0.85                 &                     \\
\hline
                            &GAIA\_DR2            &GAIA\_DR3              &SDSS\_apogee         \\
\hline
G(mag)                      &\,15.296             &\,15.272               &                     \\
$T_{eff}$(K)                &\,4,992.47           &\,5,509.6              &\,4,481.73            \\
log$g$                      &                     &\,4.51                 &\,5.0                \\
$[$Fe/H$]$                  &                     &\,-3.66                &\,-1.17              \\
distance(pc)                &\,797.64             &\,710.98               &                     \\
radius($R_{\bigodot}$)      &\,0.84               &                       &                     \\
luminosity($L_{\bigodot}$)  &\,0.399              &                       &                     \\
\hline
\end{tabular}
\end{center}

\end{table}

From the calibrated light curve, we can obtain an apparent magnitude of 12.9 for IU Leo outburst and 15.7 for quiescence, and approximately 60 day interval between outbursts (Greaves 2006). The constantly updated catalogue (Ritter \& Kolb 2003) records the orbital period of IU Leo, with the latest value of 0.376308\,days, as shown in ID\_1 of Table 2. Based on over 1,456 day span spectra observation, Thorstensen et al. (2010) derived a mass ratio of q = 0.85 for IU Leo (NSVS\,1057564+092315), marked as ID\_2 in Table 2. Pala et al. (2017) obtained an orbital period of 541.88\,minutes for IU Leo (HS 1055+0939) according the spectra from HST, marked as ID\_3 in Table 2. The orbital period of IU Leo is much longer than the period gap of 147-191 minutes (Schreiber et al. 2024) for CVs. The period gap for CVs can be explained by the disrupted magnetic braking model (Pala et al. 2020), which is observationally supported by the mass-radius relation figure of CV donor stars (Knigge 2006). On the contrary, Schaefer (2024) demonstrated the failure of the magnetic braking model by the positive rate of change in orbital period of many CVs. Theoretical models are still constantly evolving and improving. Table 2 also displays some physical data for IU Leo from GAIA\_DR2, GAIA\_DR3, and SDSS\_apogee respectively, which are helpful for comprehensive analysis. A white dwarf usually has log$g$ greater than 6.0, $T_{eff}$ hotter than 10,000\,K, and radius at a level of 0.01\,$R_{\bigodot}$. Therefore, the parameters in the lower part of Table 2 should belong to the companion star (the secondary star, Chen et al. 2022), not the white dwarf (the primary star). According to the laws of Newtonian mechanics, the mass of the companion star can be derived to be $M_{2}$ $\sim$ 0.835\,$M_{\bigodot}$ and then the mass of the primary star is $M_{1}$ $\sim$ 0.982\,$M_{\bigodot}$ based on q = 0.85. The mass and orbital period of IU Leo can be used as input parameters in the simulation calculations.

MESA is a powerful stellar evolution code that can simulate the evolution process of binary stars (Paxton et al. 2015), including both stars, double black holes, one star plus one point mass, and so on. MESA is powerful enough to handle many complex physical processes, including mass transfer from Roche-lobe overflow, gravitational wave radiation, magnetic braking, and so on. Any stellar structure and evolution code always has a lower bound of the evolutionary timescale, it is set to 10 years for MESA/Binary. It is impossible to accurately fit the outburst events within a few days. We use MESA/binary/star\_puls\_point\_mass to semi quantitatively simulate the evolution of IU Leo. The radius of the white dwarf is approximated as a point. The donor mass is set as 0.835\,$M_{\bigodot}$, the point (white dwarf) mass is set as 0.982\,$M_{\bigodot}$, and the initial orbital period is set as 0.376308\,days. Then it begins to run with the other parameters set to default. We check the evolved first model, $T_{eff}$ is 5043\,K and log$g$ is 4.62, which are basically consistent with that in Table 2. The value of luminosity is 0.322\,$L_{\bigodot}$, 19\% smaller than that in Table 2. The value of radius is 0.74\,$R_{\bigodot}$, 12\% smaller than that in Table 2. Namely, the parameters related to the companion star in Table 2 are generally self-consistent. The calculation of MESA considers the loss of angular momentum almost entirely due to gravitational radiation when the binary orbital period is smaller than the period gap. We obtain a minimum orbital period of 67 minutes, which is slightly smaller than the observed minimum orbital period of $\sim$75-82 minutes (Knigge et al. 2011).

The dwarf novae IU Leo has a white dwarf together with an accretion disk, a companion star with $T_{eff}$ $\sim$ 5,000\,K, and a circumbinary envelope with $T_{eff}$ of a few hundred Kelvin on the mid-infrared band. The white dwarf has a mass of $M_{1}$ $\sim$ 0.982\,$M_{\bigodot}$ and the companion star has a mass of $M_{2}$ $\sim$ 0.835\,$M_{\bigodot}$. With long time evolution, the magnetic braking, gravitational wave radiation, and mass loss will result in the loss of orbital angular momentum. Both the binary orbital period and the binary separation will become smaller and smaller. The total mass of the primary and companion stars will decrease. The companion star will transfer mass to the primary star. The ultimate evolutionary outcome is very complex, which depends on factors such as the size of the accretion rate and the stability of the binary system. The semi quantitative simulation helps us to have a more comprehensive understanding of the evolution process of IU Leo.

\section{The photometric analysis of IU Leo based on K2, TESS, and AAVSO}

In Sect. 3, we study the publicly available photometric data of IU Leo, investigate the binary orbital period and the outburst physical properties.

The Barbara A. Mikulski Archive for Space Telescopes (MAST) covers observations from multiple space telescopes and ground-based telescopes (Hou et al. 2023), including HST, K2, TESS, SDSS, and so on. On the MAST official website (https://mast.stsci.edu/), there are a total of 16 observations for IU Leo from HST, SDSS, K2, and TESS. We checked each observation and found 3 outbursts on the light curves, as shown in Fig. 2. For the first outburst near June 1, 2017 on the lower panel, there are only partially declined light curve records from K2. For the second outburst, we can see that the flux increased by 7.16 times, corresponding to a decrease of 2.14 apparent magnitude. For the decline process, there seems to be a standstill process, which reflects the complexity of instability in the accretion disk. For the third outburst from TESS on the upper panel, the flux increased by 5.81 times, corresponding to a decrease of 1.91 apparent magnitude. In the enlarged subgraph for the third outburst, an orbital period of IU Leo is between two minimum points on the light curves. While, for the quiescence light curves, it is a half orbital period between two minimum points on the light curves. The double minimum phenomenon of the light curve during the quiescence supports a low orbital inclination of binary stars. The brightness of the white dwarf with the accretion disk changes. The brightness will significantly increase during the outburst.

Period04 (Lenz \& Breger 2005) is a program that extracts intrinsic frequency signals from time-domain signals, which is often used to extract the pulsation frequencies of pulsating stars. For the quiescence K2 light curve in Fig. 2, we choose 82,675 data with MJD from 57916.5 to 57973.5 to derive the binary orbital period of IU Leo through Period04 program. It is 0.376307 $\pm$ 0.000004\,days. For the quiescence TESS light curve in Fig. 2, we choose 1,645 data with MJD from 59525.0 to 59536.4 to derive a binary orbital period of 0.3762 $\pm$ 0.0001\,days. The derived binary orbital period is basically consistent with the previous results shown in Table 2.

\begin{figure}
\begin{center}
\includegraphics[width=8.5cm,angle=0]{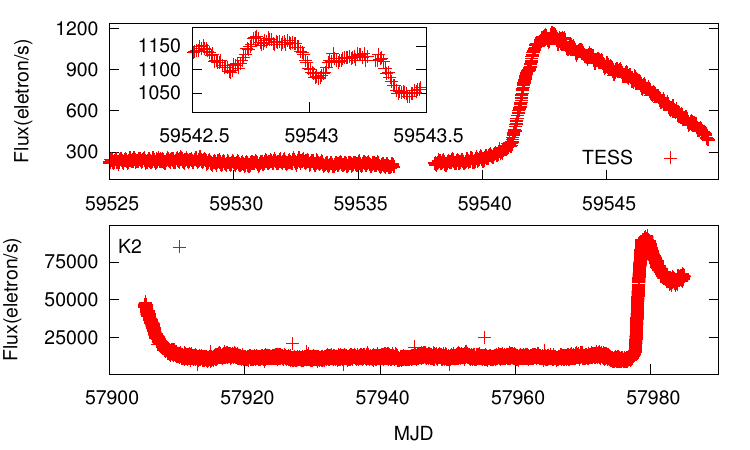}
\end{center}
\caption{High-precision light curves for IU Leo from K2 and TESS. The lower panel shows observations that began on June 1, 2017 and lasted for nearly 80 days by K2. The upper panel shows observations that began on November 7, 2021 and lasted for approximately 25 days by TESS.}
\end{figure}

\begin{figure}
\begin{center}
\includegraphics[width=8.5cm,angle=0]{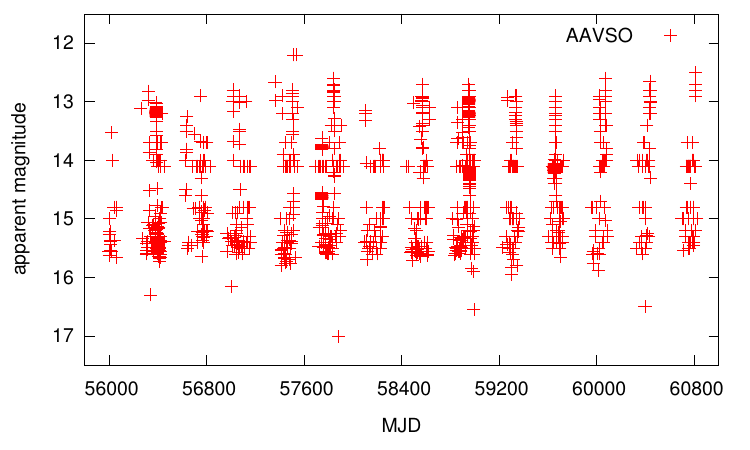}
\end{center}
\caption{The variations in apparent magnitude of IU Leo from AAVSO. The observed time is from March 13, 2012 to May 26, 2025.}
\end{figure}

\begin{figure}
\begin{center}
\includegraphics[width=8.5cm,angle=0]{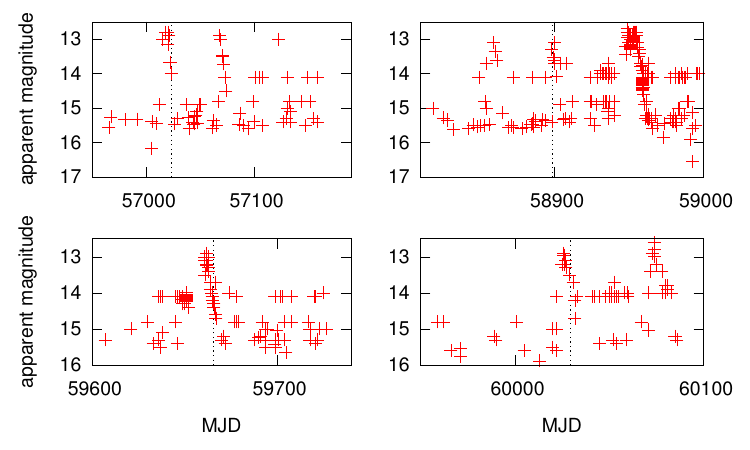}
\end{center}
\caption{The magnified figure of variations in apparent magnitude of IU Leo from AAVSO. The 4 vertical dashed lines correspond to the detailed observation times of the low resolution spectra released by LAMOST. The spectra were observed on January 1, 2015, February 19, 2020, March 27, 2022, and March 26, 2023 respectively.}
\end{figure}

There are only 3 outbursts on the light curves for the space telescope K2 and TESS. We check the variations in apparent magnitude of IU Leo from AAVSO in Fig. 3. The observed time is from March 13, 2012 to May 26, 2025. For the convenience of calculation, we approximate the observed value of "less than" a certain apparent magnitude to "equal to" the apparent magnitude. The minimum ($\sim$12.9) and maximum ($\sim$15.7) apparent magnitude  are clearly visible in Fig. 3. We checked each outburst and found that the average time interval between two outbursts was $\sim$61\,days. Both the dispersion of time interval between two outbursts and the dispersion of apparent magnitude at the brightest of each outburst reflect the instability of the accretion disk and the complexity of the accretion process.

For the LAMOST Low-Resolution Spectroscopic Survey (LRS) data release 12 (DR12), IU Leo was observed 19 times from January 1, 2015 to March 26, 2023. According to the data description, LAMOST LRS spectra cover a wavelength range from 3,700\,${\AA}$ to 9,000\,${\AA}$ with a resolution of 1,800 at 5,500\,${\AA}$. We found that there were 4 observed spectra within the outburst phase or the decline phase immediately after the outburst when we checked the specific observation times. In Fig. 4, we show a magnified figure of variations in apparent magnitude of IU Leo from AAVSO. The time interval between two outbursts are clear in Fig. 4. The detailed observation times of the 4 spectra are marked as vertical dashed lines. The spectra were observed on January 1, 2015, February 19, 2020, March 27, 2022, and March 26, 2023 respectively. Han et al. (2020) reported the decline spectra on January 1, 2015. We enriched this type of spectra to 4, as checked in Fig. 4. The identification of specific spectral lines for the 4 spectra are discussed in Sect. 4.

\section{The spectroscopic analysis of IU Leo based on LAMOST and HST}

In Fig. 5 and 6, we show 4 spectra from LAMOST LRS DR12, which corresponds to the 4 vertical dashed lines in Fig. 4. The identified emission or absorption lines are displayed in Table 3. The fluxes at the blue end of the 4 spectra are much larger than that at the red end, indicating that high-temperature accretion disk dominate the system during the outburst phase. Part of the Balmer lines have emission lines overlapping on the absorption lines, indicating that these 4 spectra are indeed in the outburst phase or the decline phase immediately after the outburst. The He II line (4,686\,${\AA}$) in upper panel of Fig. 5 is a manifestation of thermal instability in high-temperature accretion disks. We identified some metal lines from the 4 spectra, including Mg I, Na I, O I, Ca I, and Ca II. It is worth noting that there is a wide emission around 6,000\,${\AA}$ in Fig 6. The wide emission is present in all 6 spectra from LAMOST LRS DR12, which were observed on December 3, 2021, January 29, 2022, March 3, 2022, March 27, 2022, April 24, 2022, and March 26, 2023 respectively. We infer that they are instrument or reduction artefact. In fact, the circumbinary envelope in Fig. 1 is also evidence of accretion disk dynamics. For IU Leo, the Table 2 shows that the companion star are metal poor, while the Fig. 5 and 6 show that there are many metal lines in the accretion disk.

\begin{figure}
\begin{center}
\includegraphics[width=8.5cm,angle=0]{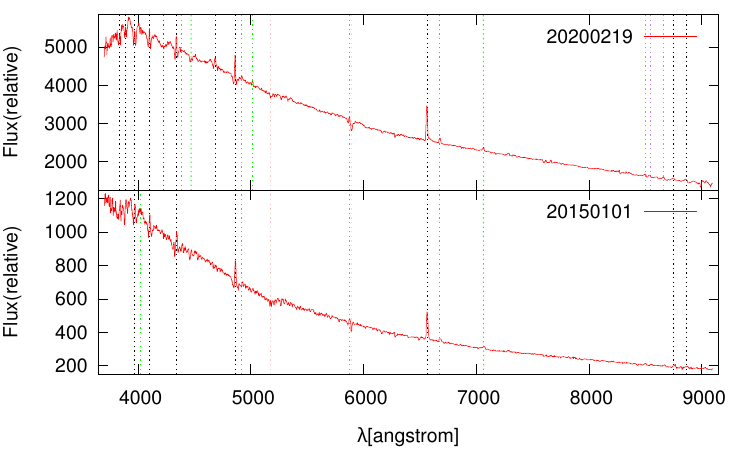}
\end{center}
\caption{The two low resolution spectra for IU Leo released by LAMOST corresponding to the vertical dashed lines in the two upper panels in Fig.\,4.}
\end{figure}

\begin{figure}
\begin{center}
\includegraphics[width=8.5cm,angle=0]{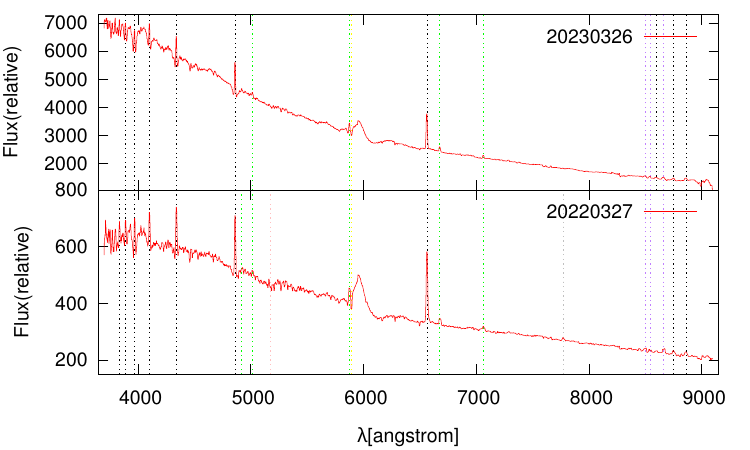}
\end{center}
\caption{The two low resolution spectra for IU Leo released by LAMOST corresponding to the vertical dashed lines in the two lower panels in Fig.\,4.}
\end{figure}

\begin{table}
\begin{center}
\caption{The wavelength of emission or absorption lines identified in LAMOST and HST.}
\begin{tabular}{llllllllllll}
\hline
lines(HST)                                                      &wavelength                       &color        \\
                                                                &(${\AA}$)                        &             \\
\hline
Ly$\alpha$                                                      &\,1215.67                        &black        \\
N V                                                             &\,1238.80,\,1242.80              &brown        \\
C II                                                            &\,1334.53,\,1335.66,\,1335.71    &blue         \\
Si IV                                                           &\,1393.76,\,1402.77              &green        \\
C IV                                                            &\,1548.20,\,1550.78              &blue         \\
He II                                                           &\,1640                           &purple       \\
\hline
lines(LAMOST)                                                   &                                 &             \\
\hline
H$\eta$                                                         &\,3836.47                        &black        \\
H$\zeta$                                                        &\,3890.15                        &black        \\
H$\epsilon$                                                     &\,3971.19                        &black        \\
H$\delta$                                                       &\,4102.89                        &black        \\
H$\gamma$                                                       &\,4341.68                        &black        \\
H$\beta$                                                        &\,4862.68                        &black        \\
H$\alpha$                                                       &\,6564.61                        &black        \\
Paschen(n=14$\rightarrow$n=3)                                   &\,8600.80                        &black        \\
Paschen(n=12$\rightarrow$n=3)                                   &\,8752.91                        &black        \\
Paschen(n=11$\rightarrow$n=3)                                   &\,8865.37                        &black        \\
He I                                                            &\,4026                           &green        \\
He I                                                            &\,4388                           &green        \\
He I                                                            &\,4472                           &green        \\
He I                                                            &\,4922                           &green        \\
He I                                                            &\,5016                           &green        \\
He I                                                            &\,5876                           &green        \\
He I                                                            &\,6678                           &green        \\
He I                                                            &\,7066                           &green        \\
He II                                                           &\,4686                           &blue         \\
Mg I                                                            &\,5176.7                         &pink         \\
Na I                                                            &\,5895.6                         &yellow       \\
O I                                                             &\,7774,\,7776,\,7777             &grey         \\
Ca I                                                            &\,4227.92                        &brown        \\
Ca II                                                           &\,8500.35,\,8544.44,\,8664.52    &purple       \\
\hline
\end{tabular}
\end{center}
\end{table}

\begin{figure}
\begin{center}
\includegraphics[width=8.5cm,angle=0]{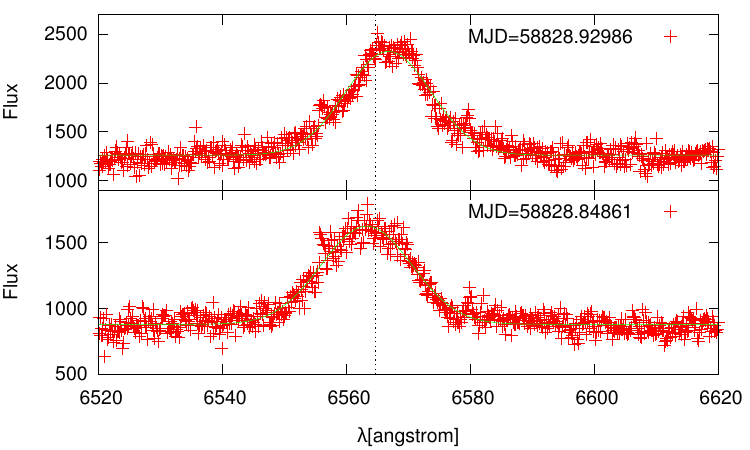}
\end{center}
\caption{The two magnified red band medium resolution spectra for IU Leo released by LAMOST MRS DR12. They were observed on December 11, 2019 and the local modified Julian minute is 84,714,012 and 84,714,129 respectively.}
\end{figure}

\begin{figure}
\begin{center}
\includegraphics[width=8.5cm,angle=0]{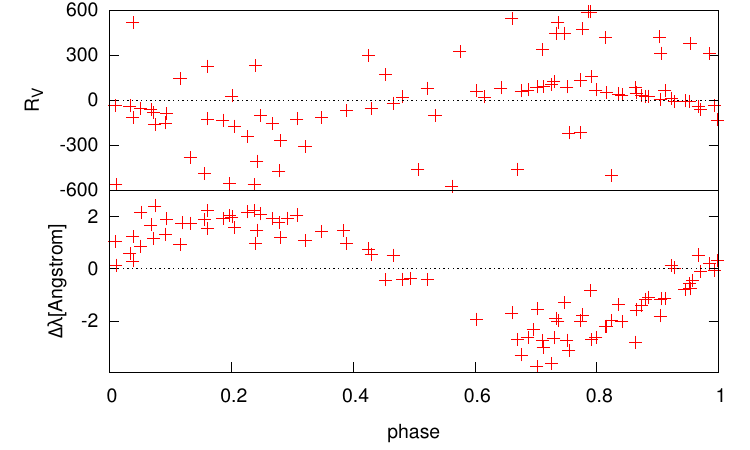}
\end{center}
\caption{The phase diagram for 90 radial velocity data and 89 maximum H$\alpha$ wavelength data from LAMOST MRS DR12.}
\end{figure}

\begin{figure}
\begin{center}
\includegraphics[width=8.5cm,angle=0]{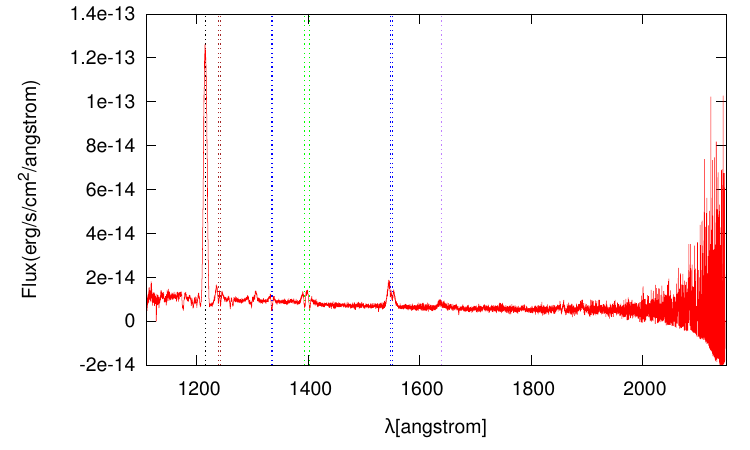}
\end{center}
\caption{The UV spectra for IU Leo released by HST from MAST.}
\end{figure}

In Fig. 7, we show two magnified red band medium resolution spectra for IU Leo released by LAMOST Medium-Resolution Spectroscopic Survey (MRS) DR12. According to the data description, LAMOST MRS spectra cover a wavelength range from 4,950\,${\AA}$ to 5,350\,${\AA}$ for blue band, and from 6,300\,${\AA}$ to 6,800\,${\AA}$ for red band with a resolution of 7,500 at 5,163\,${\AA}$ and 6,593\,${\AA}$ respectively. The LAMOST MRS DR12 released 334 spectra covering 32 Julian days from March 5, 2018 to March 3, 2023. On average, there are 5 blue band spectra and 5 red band spectra for each observed Julian day. LAMOST MRS spectra provide valuable data for studying short-term astronomical phenomena. The two spectra in Fig 7 are separated by 117 minutes with local modified Julian minute of 84,714,012 for the lower panel and 84,714,129 for the upper panel. The spectra data of wavelength from 6,520\,${\AA}$ to 6,620\,${\AA}$ are adopted and the spectra in Fig. 7 are fitted by the following equations,
\begin{equation}
Flux(\lambda)_{low} = 747.816 \times e^{-\frac{(\lambda - 6563.24)^{2}}{2 \times 7.2969^{2}}} + 878.719,
\end{equation}
\noindent
\begin{equation}
Flux(\lambda)_{up} = 1066.38 \times e^{-\frac{(\lambda - 6566.77)^{2}}{2 \times 6.9155^{2}}} + 1258.17.
\end{equation}
\noindent Then, we can derive that the velocity is -62.57\,km/s for the lower panel and 98.65\,km/s for the up panel. It indicates that the movement speed of the observed H$\alpha$ area in IU Leo changed by 161.22 km/s within 117 minutes. The CV star IU Leo indeed involves very strong dynamic processes. In Fig. 8, we show the phase diagram for total 90 radial velocity data of red band spectra for IU Leo from the official website and 89 maximum H$\alpha$ wavelength data fitted by equations like Eq\,(1) and Eq\,(2). The binary orbital period is adopted as 0.3763\,days. The lower panel of Fig. 8 shows that our method of fitting H$\alpha$ emission lines can reproduce the orbital period of binary stars.

In Fig. 9, we show the UV spectra for IU Leo released by HST from MAST. The emission lines of Ly$\alpha$, N V, C II, Si IV, C IV, and He II are identified, as shown in Table 3. The UV spectra should mainly reflect the properties of high-temperature accretion disk and white dwarf. The Ly$\alpha$ line dominates the spectra, corresponding to abundant H element. The ionization lines of N, C, Si, and He reflect different temperature gradients or regions of different energies. The UV spectra (Fig. 9) reflects the physical properties of high-temperature accretion disk and white dwarf. The optical spectra during the outburst (Fig. 5 and 6) also reflects the physical properties of high-temperature accretion disk and white dwarf. The properties of the companion star dominate the near-infrared spectrum (1.51-1.70\,$\mu$m data in SDSS\_apogee of Table 2). The properties of circumbinary envelope dominate the mid-infrared data (the WISE data of Table 1).

In Fig. 2, it seem to be a standstill phenomenon for IU Leo similar to the Z Cam subtype of CVs. However, no prolonged standstill was observed in Fig. 3. Considering an orbital period of 0.376308 days, an outburst increment of 2.4 magnitudes (Sun et al. 2021), and an average outburst interval of $\sim$61 days, IU Leo is closer to the U Gem subtype CVs, as reported by Pala et al. (2017) for IU Leo. We examined all LAMOST LRS and MRS spectra and did not find obvious double-peaked profile emission lines, as reported by Hou et al. (2020) for IU Leo. According to Fig. 1, it may be due to the strong stellar wind of the accretion disk. It may also caused by a high inclination geometry of the accretion disk.

\section{A Discussion and Conclusions}

Binary stars are crucial for studying mass transfer. The evolution of binary stars is accompanied by multi-channel angular momentum loss process. The outcome of binary star evolution strongly depends on the accretion rate and the stability of the binary star. In this paper, we conducted a comprehensive study on the dwarf nova IU Leo, a CV binary, using multi-band photometric and spectroscopic data. The data are from GAIA\_DR2, GAIA\_DR3, SDSS\_apogee, SDSS\_DR15, 2MASS, WISE, K2, TESS, AAVSO, LAMOST LRS DR12, LAMOST MRS DR12, and HST. Multi-band astronomy has prompted us to conduct more comprehensive astronomical research.

It can be inferred from the angle of IU Leo in the image provided by SDSS\_DR15 and the distance provided by GAIA that IU Leo's circumbinary envelope extends to approximately $\sim$3,745\,AU on the optical band, which is obviously larger than the binary separation (a few $R_{\bigodot}$). The apparent magnitude values for IU Leo from SDSS\_DR15, 2MASS, and WISE correspond to a low-temperature blackbody radiation with a few hundred Kelvin ($\sim$132\,K) on the mid-infrared band. It is the temperature of the circumbinary envelope matter. Because the temperature of the primary star white dwarf is usually above 10,000\,K and the temperature of the companion star had been measured to be $\sim$5,000\,K from GAIA\_DR2, GAIA\_DR3, and SDSS\_apogee. The low-temperature circumbinary envelope of IU Leo dominates the mid-infrared data. Hoard et al. (2009) reported a maximum inner edge temperature of $\sim$500\,K for circumbinary cool dust of novalike cataclysmic variable V592 Cassiopeiae based on the mid-infrared band observation data released by the Spitzer Space Telescope. The companion star of IU Leo dominates the near-infrared spectra. While, the high-temperature accretion disk and primary white dwarf of IU Leo determine the UV spectra and outburst optical spectra.

The physical parameters of IU Leo are reviewed in Table 2. With the module of MESA/binary/star\_puls\_point\_mass, we adopt $M_{2}$\,=\,0.835\,$M_{\bigodot}$, $M_{1}$\,=\,0.982\,$M_{\bigodot}$, and an initial orbital period of 0.376308 days to semi quantitatively simulate the evolution of IU Leo. The evolution results are basically consistent with the parameters in Table 2, such as the effective temperature, the gravitational acceleration, the stellar radius, and the stellar luminosity. This evolutionary results indicate that the physical parameters in Table 2 are basically self consistent. Zorotovic et al. (2011) reported that a large white dwarf mass in CVs was common and an average white dwarf mass in CVs was 0.83 $\pm$ 0.23\,$M_{\bigodot}$. The simulated minimum orbital period is 67\,minutes, slightly smaller than the observed minimum orbital period of $\sim$75-82 minutes (Knigge et al. 2011).

We examined high-precision data for IU Leo from K2 and TESS and found 3 outbursts. When IU Leo bursts, its brightness increases by $\sim$2 magnitudes. The orbital period of 0.3763 days is clearly visible on the outburst light curves. For the quiescence K2 light curve and TESS light curve in Fig. 2, we derived a binary orbital period of 0.376307 $\pm$ 0.000004\,days and 0.3762 $\pm$ 0.0001\,days respectively by the Period04 program. They are consistent with the previous results of Ritter \& Kolb (2003), Thorstensen et al. (2010), and Pala et al. (2017). By comparing with AAVSO data, we identified 4 spectra with outburst phase or just completed outburst from LAMOST LRS DR12. Three of them are first reported. Partial Balmer emission lines are clearly visible when overlapping on the absorption lines. The average regular outburst of $\sim$61 days (AAVSO data) comes from the instability of the accretion disk. Based on the red band spectra of LAMOST MRS DR12, we fitted a total of 89 maximum H$\alpha$ wavelength data and obtained a phase diagram for IU Leo, as shown in Fig. 8. The binary orbital period is set as 0.3763\,days. Fitting the typical emission lines of the spectra of LAMOST MRS data can detect the binary orbital period. The spectroscopy is very useful for studying astrophysics. In addition, we have identified many H, He, C, N, O, Na, Mg, Si, and Ca neutral and ionized lines among spectra of IU Leo. As early as 1980, Williams reported that the Balmer lines and Ca II and K emission in quiescence of CVs were from the optically thin outer regions of the disks, the thermal continuum spectrum was emitted by the inner disk, and the He I and He II lines were produced by different mechanism.

The CVs contains rapid dynamic processes and rich physical laws, making it a natural laboratory for studying physical laws. The determination of basic physical parameters, research on angular momentum loss mechanisms, and study of emission line excitation mechanisms are all meaningful research hotspots for CVs. The use of spectra released by LAMOST MRS to explore the orbital periods of CVs provides an effective entry point for preliminary research. In the future, we will use the large amount of LAMOST MRS data, together with multi-band photometric and spectroscopic data, to conduct more comprehensive and in-depth research on CVs.

\section{Acknowledgment}

Guoshoujing Telescope (the Large Sky Area Multi-Object Fiber Spectroscopic Telescope LAMOST) is a National Major Scientific Project built by the Chinese Academy of Sciences. Funding for the project has been provided by the National Development and Reform Commission. LAMOST is operated and managed by the National Astronomical Observatories, Chinese Academy of Sciences. The work is supported by the Yunnan Provincial Department of Education Science Research Fund Project (No. 2024J0964) and the International Centre of Supernovae, Yunnan Key Laboratory (No. 202302AN36000101).

\label{lastpage}

\end{document}